\documentclass[twocolumn,showpacs,preprintnumbers,amsmath,amssymb]{revtex4-1}
\usepackage{graphicx}% Include figure files
\usepackage{dcolumn}% Align table columns on decimal point
\usepackage{bm}% bold math
\usepackage{color}

\usepackage{hyperref}
\usepackage{amssymb}
\usepackage{amsmath}
\usepackage{amsfonts}
\usepackage{parskip}
\usepackage{empheq}

\newcommand{\beq}{\begin{equation}}
\newcommand{\eeq}{\end{equation}}
\newcommand{\bea}{\begin{eqnarray}}
\newcommand{\eea}{\end{eqnarray}}

\newcommand{\comment}[1]{}

\begin{document}

%\preprint{}

\title{Large-scale Ising spin network based on degenerate optical parametric oscillators}

\author{Takahiro Inagaki$^{1}$} %\email{sasaki@qi.t.u-tokyo.ac.jp}
\author{Kensuke Inaba$^{1}$}
\author{Ryan Hamerly$^{2}$}
\author{Kyo Inoue$^{3}$} 
\author{Yoshihisa Yamamoto$^{2}$}
\author{Hiroki Takesue$^{1}$} \email{takesue.hiroki@lab.ntt.co.jp}
\affiliation{%
$^1$NTT Basic Research Laboratories, NTT Corporation, 3-1 Morinosato Wakamiya, Atsugi, Kanagawa, 243-0198, Japan\\
$^2$E. L. Ginzton Laboratory, Stanford University, Stanford, CA94305, USA\\
$^3$Division of Electrical, Electronic and Information Engineering, Osaka University, Osaka 565-0871, Japan\\
}%

\date{\today}% It is always \today, today,

\if0
\begin{abstract}

\end{abstract}
\fi

%\pacs{42.65.Ky, 42.50.Dv, 03.67.Hk}% PACS, the Physics and Astronomy
\maketitle

%1st paragraph
%{\bf bold font here.}

%\section{Introduction}

{\bf Simulating a network of Ising spins with physical systems is now emerging as a promising approach for solving mathematically intractable problems \cite{dwave,ion,utsunomiya,cmos,imran}.
% that are hard to compute with modern computers. 
%Recently, it has been proposed that the ground-state-search problem of the Ising Hamiltonian can be solved with a network of laser oscillators \cite{utsunomiya,wang,haribara,alireza}. 
%, and a proof-of-principle experiment has been reported using a small system consists of four signal/idler degenerate optical parametric oscillators (DOPO) \cite{alireza}.   
Here we report a large-scale network of artificial spins based on degenerate optical parametric oscillators (DOPO), paving the way towards a photonic Ising machine capable of solving difficult combinatorial optimization problems. We generated $>$10,000 time-division-multiplexed DOPOs using dual-pump four-wave mixing (FWM) \cite{mck,fan} in a highly nonlinear fibre (HNLF) placed in a fibre cavity. Using those DOPOs, a one-dimensional (1D) Ising model was simulated by introducing nearest-neighbour optical coupling. 
We observed the formation of spin domains and found that the domain size diverged near the DOPO threshold, which suggests that the DOPO network can simulate the behaviour of low-temperature Ising spins. 
% which suggests that the DOPO network can find its ground state within a power-law time-scaling of the network size. 
%We confirmed the ferro- and anti-ferromagnetic behaviours of DOPOs with formation of domains, with which we could observe the emergence of a long-range spin order near the threshold.  
%The present system is very simple, but it is hard to solve because the 1D nature in the large scale system enables the formation of the stable domain-walls.
%The present results obtained with the 1D system composed of 10,000 spins suggested the expected computational power.
 }

Combinatorial optimization problems are becoming increasingly important in our society, for example in applications such as artificial intelligence, drug discovery, optimization of cognitive wireless networks, and analysis of social networks. Many such problems are classified as non-deterministic polynomial time (NP)-hard or NP-complete problems, which are considered to be hard to solve efficiently with modern computers \cite{combi}. 
It is well known that many combinatorial optimization problems can be mapped onto the ground-state-search problems of the Ising Hamiltonian \cite{ising}. Various schemes have been proposed and demonstrated that simulate the Ising Hamiltonian with physical systems, such as superconducting circuits \cite{dwave}, trapped ions \cite{ion}, CMOS devices \cite{cmos}, and electro-mechanical oscillators \cite{imran}. Among such schemes, a coherent Ising machine (CIM) is now attracting attention \cite{utsunomiya}. A CIM simulates the Ising model using a network of lasers with binary oscillation conditions as artificial spins, and is expected to have a significant advantage in terms of computation time over conventional schemes such as simulated annealing and semi-definite programming \cite{utsunomiya,takata,wang,haribara}. Recently, Marandi et al. demonstrated a CIM using DOPOs \cite{alireza}. A DOPO can be utilized as a stable artificial spin because it takes only the 0 or $\pi$ phase relative to the pump phase \cite{nabor}. The spin-spin interaction can be simply implemented with mutual injections of DOPO lights using delay interferometers. In \cite{alireza}, a spin system composed of four DOPOs was employed for a proof-of-principle CIM experiment. However, to simulate a more complex Ising Hamiltonian to verify the advantages of the CIM over existing methods, we need to implement a CIM with a much larger number of spins. 
Here we report a large scale network of artificial spins realized with as many as 10,000 time-division-multiplexed DOPOs generated via dual-pump FWM in an HNLF placed in a fibre cavity. We successfully simulated the ferro- and anti-ferromagnetic-like behaviour of a 1D Ising spin chain by introducing uni-directional nearest-neighbour coupling between DOPOs. In addition, we observed a formation of domain walls with which we could obtain information on how much the state of the spin network was excited from the ground state. We believe the present result will provide a promising platform on which to realize an efficient machine for solving the Ising model based on the CIM concept. 

%\if0
\begin{figure*}[th]
\centerline{\includegraphics[width=.9\linewidth]{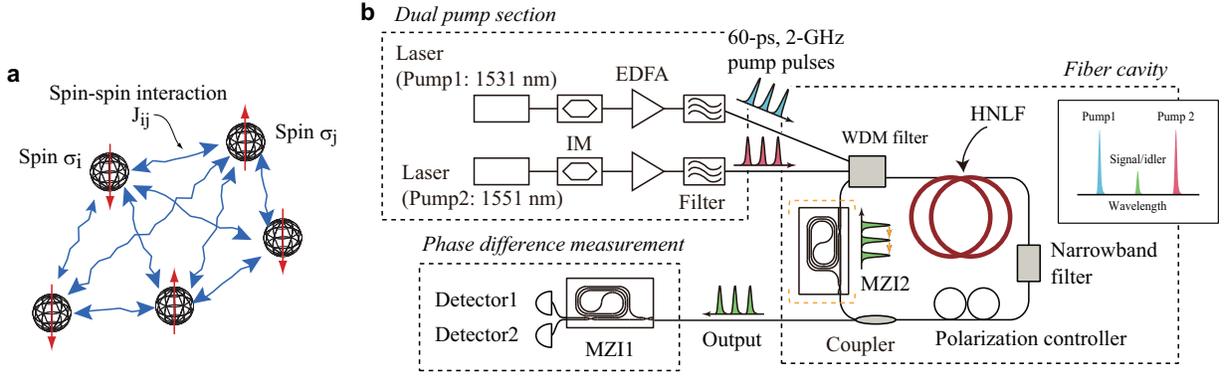}}
\caption{Ising model and setup for generating artificial Ising spins based on DOPOs. {\bf a} An Ising model. {\bf b} Experimental setup. IM: intensity modulator, EDFA: erbium doped fibre amplifier, WDM: wavelength division multiplexing, HNLF: highly nonlinear fibre, MZI: delayed Mach Zehnder interferometers. The difference between the propagation times of the two arms of MZIs are 500 ps for both MZI1 and 2. The inset shows the wavelength allocation of pump 1, 2 and the signal/idler wave. MZI2 is inserted when simulating 1D Ising model.  }
\label{setup}
\end{figure*}
%\fi

%\section*{Results}

%\subsection*{Coherent Ising machine}
A dimensionless Hamiltonian of an $N$-spin Ising model without an external magnetic field (Fig. \ref{setup} {\bf a}) is given by
\begin{equation}
H=-\sum_{1 \le i<j \le N} J_{ij} \sigma_i \sigma_j, \label{hami}
\end{equation}
where $J_{ij}$ is the coupling coefficient between the $i$th and $j$th spins, and $\sigma_\ell$ $(\ell\in \{i,j\})$ denotes the $z$ projection of the $\ell$th spin, which can take $\pm 1$ values. The purpose of an Ising machine is to find the ground state of the above Hamiltonian with a given set of $J_{ij}$ using a physical system. To realize an Ising machine, we need elements with a binary degree of freedom to represent the spins and a method for realizing programmable coupling between spins.
In a CIM, we employ two-mode laser oscillators \cite{utsunomiya,takata} or DOPOs \cite{wang,haribara,alireza} as artificial spins. The spin coupling can be implemented by injecting a portion of light from the $i$th spin into the $j$th spin and vice versa. Therefore, we can set $J_{ij}$ by changing the phase and transmittance of the optical paths that connect the $i$th and $j$th spins. For CIM operation, we start with a zero pump power for all the oscillators, and set the $J_{ij}$ values by establishing optical paths between the spins. We then gradually increase the pump. 
With $N$ spins, there are $2^N$ combinations of spin configurations. In other words, we are operating a multi-mode oscillator with $2^N$ modes. 
As we increase the pump, the network reaches the threshold, and an oscillation starts most likely at the mode (or spin configuration) with the lowest loss, which will give the ground state of the Ising Hamiltonian.   
 
%\subsection*{Phase sensitive amplification with dual-pump FWM}
A DOPO can be realized by placing a phase sensitive amplifier (PSA) \cite{mck,fan,marhic,levenson,choi,imajuku,tong} in a cavity, where only a signal with phase 0 or $\pi$ relative to the phase of the pump for parametric amplification process is amplified.  
The principle of PSA with dual-pump FWM is detailed in Method. 
When we install a PSA in a cavity and drive it with a below-threshold pump, we observe a quadrature-squeezed noise generated by spontaneous parametric downconversion or spontaneous FWM. As we increase the pump power, the noise light undergoes phase sensitive amplification, which leads to phase bifurcation as a result of spontaneous symmetry breaking \cite{alireza,nabor,alireza2}. When the pump power reaches the threshold, we obtain DOPOs whose phases can take only 0 or $\pi$. Since the oscillation is initiated with the noise photons generated by spontaneous parametric processes, the emergence probabilities of $0$ and $\pi$ phases are inherently equal. 
If we employ pulsed pump with a temporal separation $\Delta t$, we can generate $N$ independent DOPOs with a single cavity by satisfying a condition $T_c = N \Delta t$, where $T_c$ denotes the cavity round-trip time. The characteristics of these DOPOs are essentially identical except for their phases, since they share the same cavity. Thus, we can increase the number of spins simply by increasing the pump repetition frequency or by increasing $T_c$.  
%In the experiment reported in \cite{alireza}, a PSA based on parametric downconversion in a periodically poled lithium niobate crystal was used in a few-meter free-space cavity to generate four time-division-multiplexed DOPOs. In our experiment, we employ a PSA based on dual-pump, signal-idler degenerate FWM \cite{mck,fan} in a 1-km HNLF in a fibre cavity.
%, pumped at a 2-GHz repetition frequency. 

%\if0
\begin{figure*}[thb]
\centerline{\includegraphics[width=.9\linewidth]{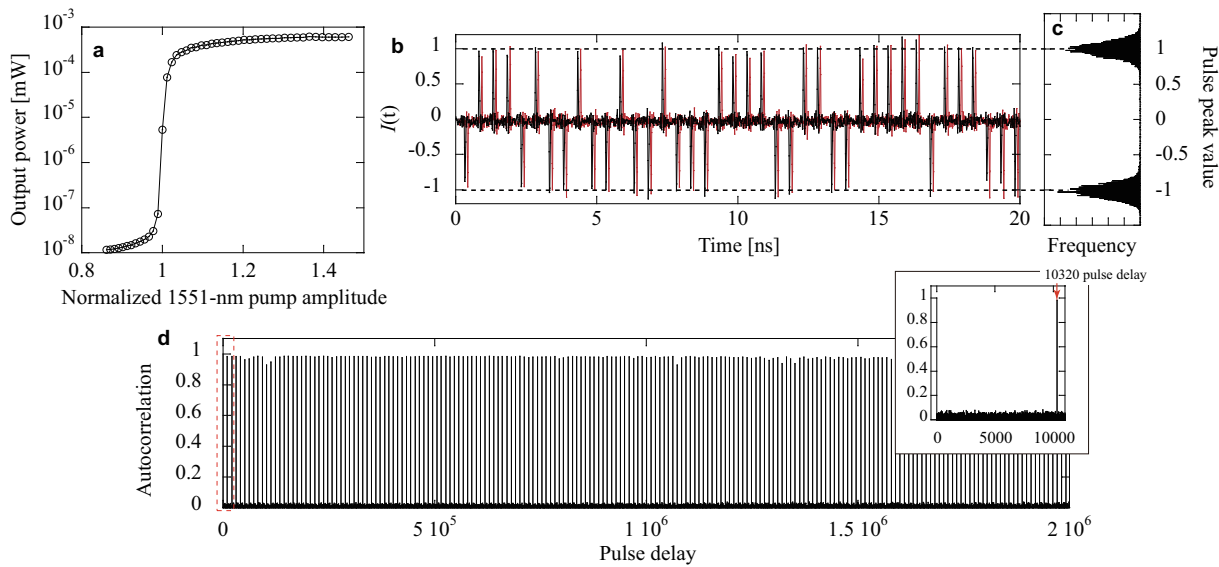}}
\caption{DOPO measurement results (without optical coupling). {\bf a} DOPO output power as a function of normalized 1551-nm pump amplitude. {\bf b} Temporal waveforms of the phase measurement signal $I(t)$. The red curve shows a waveform whose temporal position was shifted by $10320\Delta t$. To distinguish the no-shift waveform shown by the black curve, we inserted a 100-ps offset between the black and red curves.  {\bf c} Histogram of the pulse peak values for 10320 DOPOs. Clear phase discretization is observed. {\bf d} Autocorrelation measurement result. The inset shows the magnification of the area shown by the red dotted square. }
\label{wfm}
\end{figure*}
%\fi

%\subsection*{$>$10,000 DOPO generation}

Figure \ref{setup} shows the experimental setup. Continuous waves from two lasers with wavelengths of 1531 and 1551 nm were modulated into 2-GHz, $\sim$100 ps pulse trains using lithium niobate intensity modulators. The pulse trains were amplified by erbium-doped fibre amplifiers (EDFA) and passed through optical bandpass filters to suppress the amplified spontaneous emission noise from the EDFAs. The amplified pulse trains were injected into a fibre cavity through a wavelength division multiplexing (WDM) filter. The fibre cavity contained a 1-km HNLF, an optical bandpass filter whose passband width was 25 GHz, a polarization controller, a 99:1 coupler for extracting a portion of the OPO light, and the WDM filter. The HNLF had a zero dispersion wavelength of 1537 nm and a nonlinear coefficient of 21 [/W/km] (specification). The centre wavelength of the optical bandpass filter was set at 1541 nm so that only light that satisfied the signal-idler degenerate condition could oscillate. We obtained phase sensitive amplification through dual-pump FWM in the HNLF. As we increased the pump powers to exceed the threshold, we obtained a group of DOPOs. Since the cavity round-trip time $T_c$ was approximately 5.2 $\mu$s and the pump pulse interval $\Delta t$ was 500 ps, we could generate $>$10,000 DOPOs multiplexed in the time domain. The DOPO train extracted from the 99:1 coupler was launched into a 1-bit delay Mach-Zehnder interferometer (MZI1) whose two outputs were each connected to a photodetector. The phase difference between the two arms of the interferometer was adjusted so that the light was detected by detector 1 (2) if the phase difference between adjacent DOPOs was 0 ($\pi$). Hereafter, the phase difference measurement result is represented by $I(t)=I_1(t)-I_2(t)$, where $I_1(t)$ and $I_2(t)$, respectively, correspond to the normalized photocurrents at time $t$ observed with detectors 1 and 2. Note that the peaks in the waveform of $I(t)$ represent $\cos \Delta \theta_i$, where $i$ and $\Delta \theta_i$ denote the index of a DOPO and the phase difference between the $i$th and $(i-1)$th DOPOs, respectively.   
%This means that the phase differences 0 and $\pi$ are observed as $\Delta I(t)=1$ and $-1$, respectively.  

Figure \ref{wfm} {\bf a} shows the OPO output power as a function of the 1551-nm pump amplitude normalized by that at the threshold ($\sim$ 10-mW peak power), which we denote by $p$ hereafter. Here we fixed the 1531-nm pump peak power at $\sim$0.5 W. Thus, we observed that the OPO output power exhibited clear threshold behaviour. 
%The OPO power inside the cavity was around $10^{-7}$ at the threshold. 
Figure \ref{wfm} {\bf b} shows a result of phase difference measurement $I(t)$. 
The sign of $I(t)$ changed randomly for each pulse, while the amplitude remained almost the same. 
The measured pulse width was $\sim 30$ ps. 
Figure \ref{wfm} {\bf c} shows a histogram of the peak values of $I(t)$, which clearly indicates the discretization of the DOPO phase into 0 and $\pi$. The ratio between the positive and negative pulses was 1:0.996, indicating that the probabilities of the emergence of 0 and $\pi$ were the same. 
%Thus, the phase difference between pulses were discretized to $\{0,\pi\}$, as expected for DOPOs. 
%The slight fluctuation of the amplitudes is probably due to the instability of the setup, presumably caused by the frequency noises of the pump lasers.  
%Although a slight fluctuation of pulse powers are seen, we observed a clear discretization of the amplitude of $\Delta I$, which clearly suggests the phase difference between pulses were bifurcated to $\{0,\pi\}$.
% where complementary binary bit patterns were obtained at detector 1 and 2. 
%This means that the phase of pulses were bifurcated to $\{0,\pi\}$, which is a clear sign of DOPOs. 

Although the phases of $N$ DOPOs generated in our setup are completely random, each OPO should preserve the phase once the pump power exceeds the threshold level. 
This means that the same random pattern should be observed in the phase difference measurement for every $N \Delta t$. To confirm this, we took the phase difference measurement result for 2,000 pulses and calculated the auto-correlation (see Supplementary Information). % defined by the following equation as a function of pulse delay $\Delta i$. 
%\begin{equation}
%C(\Delta i) =\frac{\sum_{i=1}^n(I(i)-\overline{I(i)})(I(i+\Delta i)-\overline{I(i)})}{\sqrt{\sum_{i=1}^n(I(i)-\overline{I(i)})^2}\sqrt{\sum_{i=1}^n(I(i+\Delta i)-\overline{I(i)})^2}}
%\end{equation}
%Here, $I(i)$ denotes the discretized phase difference measurement result at the $i$th pulse and $n=2000$. 
The result is shown in Fig. \ref{wfm} {\bf d}, and the region around 0 pulse delay is enlarged in the inset. As seen, an identical phase pattern was obtained 93 times for every 10,320 pulses, which means that each DOPO was oscillating with the same phase for at least 93 circulations in the cavity. 
The red curve in Fig. \ref{wfm} {\bf b} shows the $I(t)$ waveform shifted by 10,320 pulses (with a 100 ps offset for clarity). Thus, we could confirm that an identical phase pattern was repeated after a circulation of the pulse train. 
These results indicate that our DOPOs could be operated stably at well above the threshold. 

%$\Delta \theta_i$は$\cos \Delta \theta_i \simeq (I_1(i)-I_2(i))/(I_1(i)+I_2(i))$と得られる、E

%\subsection*{Simulation of one dimensional Ising model with DOPOs}

%\if0
\begin{figure}[thb]
\centerline{\includegraphics[width=.9\linewidth]{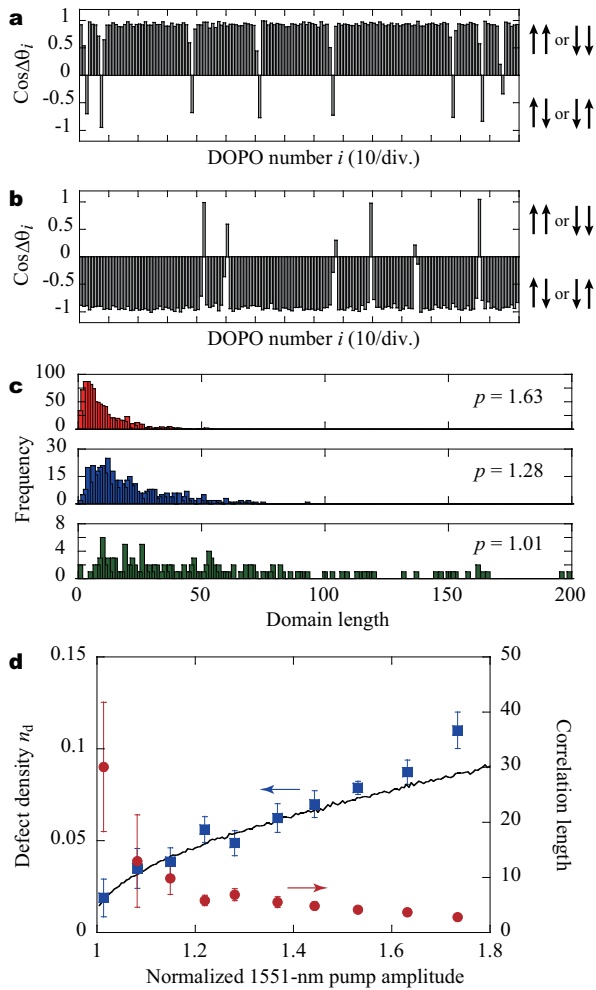}}
\caption{Results observed with $>$10,000-spin 1D Ising machine. Example phase difference measurement results for {\bf a} coupling phase 0 and {\bf b} $\pi$ for a normalized 1531-nm pump amplitude of 1.40 {\bf c} Histograms of domain length distributions for $p=1.63$ (red columns), 1.28 (blue), and 1.01 (green). {\bf d} Defect density $n_{\rm d}$ and correlation length $x_0$ as a function of normalized 1551-nm pump amplitude. Squares: $n_d$ (experimental data), solid line: $n_d$ (numerical simulation), circles: $x_0$ (experimental data). The experimental data were the average of the values obtained from phase difference measurements performed five times at each pump amplitude.   }
\label{1d}
\end{figure}
%\fi

We realized a simulator of a 1D Ising model by inserting a 1-bit delay interferometer (MZI2) that was similar to MZI1 into the fibre cavity (Fig. \ref{setup}). The function of MZI2 was to extract half  of the power  of the $i$th pulse and inject it into the $(i+1)$th pulse for $i<N$, and a portion of the $N$th pulse was launched into the 1st pulse. 
This means that with this setup we simulated the Hamiltonian given by 
$H = - \sum_{i=1}^N J \sigma_i \sigma_{i+1}$, 
with a periodic boundary condition $\sigma_{N+1}=\sigma_1$, which corresponds to the 1D Ising Hamiltonian analysed in Ising's original paper \cite{ising2}.  
%With the MZI2 that consisted of two 3-dB couplers, the absolute value of the coupling coefficient $|J|$ can be approximated to be 0.5 \cite{ryan}. 
%This means that we implemented uni-directional nearest-neighbour coupling between DOPOs, namely $J_{i \to i+1}=\pm 1$ and $J_{i+1 \to i}=0$. 
The sign of $J$ can be changed by tuning the phase of the delayed interferometer: sgn$(J)=1$ and $-1$ can be realized by setting the phase difference of the interferometer at 0 and $\pi$, respectively. We measured the phase difference of the DOPOs from the cavity with the setup described in the previous section. $N$ was increased to 10337 in this experiment, which was due to the increase in the fibre cavity length caused by the insertion of MZI2.

The phase difference between adjacent DOPOs, $\cos \Delta \theta_i$, for the coupling phase 0 and $\pi$ are shown in Fig. \ref{1d} {\bf a} and {\bf b}, respectively, at a normalized 1551-nm pump amplitude of 1.40. 
%In the following, $I(i)$ denotes the result of the phase difference measurement between the $i$th and $(i-1)$th DOPOs.  
When the coupling phase was set at 0, $\cos \Delta \theta_i$ was $\sim 1$ for the majority of pulses, implying that the phases of the DOPOs were now aligned so that they were in phase. 
 With the $\pi$ coupling phase, $\cos \Delta \theta_i$ was mostly negative, which means that the adjacent pulses now had alternating phases. 
 These phase configurations are analogous to ferromagnetic and anti-ferromagnetic spin configurations, respectively. 
 We also measured the phase difference as we changed the coupling phase. 
We observed a sharp transition from the ferromagnetic to the anti-ferromagnetic spin configuration at a coupling phase of $(2k+1) \pi/2$ ($k$: integer), which confirmed that the DOPO phase was discretised even at the boundary between 0 and $\pi$ phase coupling (see Supplementary Information for details).  

It is well known that no phase transition occurs in a 1D Ising model at finite temperatures \cite{ising2,domain}. This means that, when $N$ is large, all the spins are not aligned in the same value in a 1D Ising model, and instead we observe the formation of stable domains at a temperature greater than absolute zero. 
Interestingly, we observed several inverted peaks in the $\cos \Delta \theta_i$ measurement results in both Fig. \ref{1d} {\bf a} and {\bf b}. This suggests that we observed in- and anti-phase ``domains" in Fig. \ref{1d} {\bf a} and {\bf b}, respectively, and the inverted peaks correspond to the boundaries of domains (domain walls). 
%The blue columns in Fig. \ref{1d} (c) shows the domain length distribution histogram for data shown in Fig. \ref{1d} (a). This histogram suggests that the spin flip is occurred in a probabilistic way. 
%and thus we can reasonably assume that the probability of finding a spin in an excited state obeys Boltzmann distribution. 
The domain length distribution histograms for various pump amplitudes are shown in Fig. \ref{1d} {\bf c}, which clearly suggests that the interaction length between spins became longer as the pump amplitude was set closer to the threshold. 

We can estimate the energy increase of the spins from the ground state by counting the number of domain walls, which we refer to as $N_{\rm d}$ hereafter.  
In our 1D Ising system, one spin flip from the ground state increases the total energy by $2 J$.  Therefore, the energy increase per spin from the ground state is given by $2 J n_{\rm d}$, where $n_{\rm d}$ is the defect density $N_{\rm d}/N$. 
The experimentally obtained defect densities $n_{\rm d}$ are plotted as a function of pump amplitude in Fig. \ref{1d} {\bf d}. 
The result agrees well with a numerical simulation based on the discrete-time model described in Method, which is shown by the solid line. 
%This result confirmed that the spin energy decreases as the pump amplitude were set closer to the threshold, which suggests that the CIM would perform best when we pump the DOPOs near the threshold. 

We also took the auto-correlation of the phase difference measurement data for various $p$ values, and fitted it with a function $\exp\left(-x/x_0\right)$, where $x_0$ denotes the correlation length. The obtained $x_0$ values as a function of $p$ are shown by the circles in Fig. \ref{1d} {\bf d}. 
%The correlation length diverged when $p \sim 1$, implying that a longer range order can be formed among the spins as $p$ approaches the threshold value. \textcolor{red}{This result suggests that the CIM can in principle find a ground state of the spin system when .. (please state something about computation). }
The correlation length diverged as $x_0 \propto 1/(p-1)$, implying that a longer range order can be formed when $p$ approaches 1.
On the other hand, it takes a longer time to reach the DOPO transitions when $p$ approaches 1.
The time to reach the DOPO transitions, which we call the saturation time $T_s$, is analytically related to $x_0$ as $T_s \propto
x_0^2$ (see Method). 
Our results show that the ferro- or anti-ferromagnetic ground states of the 1D
Ising model realized with the DOPO can in principle find the ground state of the $N$-spin systems ($N<x_0$)
within a power-law time-scaling of $N^2$.
Note that the 1D Ising model is in fact a hard problem to compute with a physical system because a 1D spin system suffers from larger spin fluctuations than those in a higher dimensional spin system. 
% , and thus investigating the behaviour of a large-$N$ system with a higher dimensional coupling will constitute an important future work to evaluate the computational power of the CIM.  }
%, and thus the testbeds in the huge-$N$ systems are important to evaluate the computational power in future.

It is informative to estimate the normalized temperature $T_n := k_B T/J$ of the spin system, where $T$ and $k_B$ denote absolute temperature and the Boltzmann constant, respectively. The defect density $n_d$ and the correlation length $x_0$ can be related to $T_n$ with the following equations: $2 n_d = 1-\tanh\left(1/T_n\right)$, $x_0 = -1/\ln \left(\tanh\left(1/T_n\right)\right)$. 
%From the histogram shown in Fig. \ref{1d} {\bf c}, it is clear that our spin system is not exactly at thermal equilibrium: the domain length histogram should follow an exponential distribution at thermal equilibrium, while that for our system deviated from an exponential distribution at small domain lengths. Still, it is informative to approximate our spin system is at thermal equilibrium so that we can relate the defect density to the effective temperature of our spins with the following equation: $-\tanh \left(\frac{J}{T} \right) = -1 + 2 n_{\rm d}$. 
For example, at a normalized pump amplitude of 1.01 and with in-phase coupling, both of the above equations consistently gave $T_n \simeq 0.5$. 

These results obtained from observation of the 1D Ising model show that our DOPOs well simulated the behaviour of a low-temperature spin system. 
We expect that the normalized temperature of a 1D Ising system can be a useful index with which to evaluate the quality of both DOPOs and other physical systems that constitute Ising machines. 
%An important consideration in developing future Ising systems may be the degree to which we can further ``cool'' the spins. 
How much further we can``cool'' the spins may be an important consideration in developing Ising systems in the future.

\section*{Author contributions}
T. I. and H. T. constructed the DOPO setup and performed the experiments. R. H. and K. Inaba developed the theoretical model. T. I., K. Inaba, R. H. and H. T. analysed the data. 
%R. H, K. Inaba. and H. T. undertook numerical analysis. 
H. T., K. Inoue. and Y. Y. conceived the concept of the experiment. All the authors discussed the results and wrote the paper.   

\section*{Acknowledgements}
The authors thank Shoko Utsunomiya, Alireza Marandi, Peter McMahon, Koji Igarashi, Shuhei Tamate, Kenta Takata, Yoshitaka Haribara, and Kaoru Shimizu for fruitful discussions. This research was funded by the ImPACT Program of the Council of Science, Technology and Innovation (Cabinet Office, Government of Japan).

\section*{Methods}

\noindent
{\bf Discrete-time model for simulating DOPO based on dual-pump FWM. }
Since the DOPOs in this experiment operated with a high gain because of the relatively large cavity loss, the continuous-time model for simulating DOPOs reported in \cite{wang,haribara} does not accurately simulate the present experiment and so we needed to employ a discrete-time model. Here we briefly describe the discrete-time simulation of the DOPOs. This model will be reported in detail elsewhere. 
 
We assume that the complex amplitudes of a degenerate signal, and two pumps are represented by $a$, $b$ and $c$. When the phase-matching condition is satisfied, the mode coupling equations for these amplitudes are expressed as \cite{agrawal,yyki} 
%When an optical pulse propagates down a fibre, it obeys a set of field equations that account for dispersion, self- and cross-phase modulation and four-wave mixing \cite{agrawal}.  Since the dispersion length is longer than the fibre in this case, dispersion can be ignored.  The self- and cross-phase modulation terms are all either negligible or cancelled with phase matching.  The only nonlinearity we need to consider, therefore, is four-wave mixing.  For a degenerate signal $a$ and pumps $b$, $c$ (with $|b| \ll |c|$), the field equations become:

%\textcolor{blue}{
\bea
	\frac{da}{dz} & = & 2i \gamma a^* b c - \frac{1}{2}\alpha a \\
	\frac{db}{dz} & = & -i \gamma a^2 c^* - \frac{1}{2}\alpha b \\
	\frac{dc}{dz} & = & -i \gamma a^2 b^* - \frac{1}{2}\alpha c		
\eea
%}

where $z$ is the position along the HNLF and $\gamma$ is the nonlinear coefficient of the HNLF.  
%We may assume $b$ and $c$ are real without loss of generality, and since we are interested in above-threshold behaviour, $a$ may be presumed real because only the real quadrature of $a$ experiences gain.  
%We express the phase terms of $a$, $b$ and $c$ with $\phi_a$, $\phi_b$ and $\phi_c$. 
We rescale the field amplitudes as: $x = (2\gamma L_{\rm eff})^{-1/2} e^{-\alpha z/2}\bar{x}$ ($x \in {a,b,c}$), where ($L_{\rm eff} = (1 - e^{-\alpha L})/\alpha$), and of the distance $s = (1 - e^{-\alpha z})/(1-e^{-\alpha L})$. Then we obtain the following normalized equations.

%\textcolor{blue}{
\beq
	\frac{d\bar{a}}{ds} = 2i\bar{a}^*\bar{b}\bar{c},\ \ \ 
	\frac{d\bar{b}}{ds} = -i \bar{a}^2 \bar{c}^*,\ \ \ 
	\frac{d\bar{c}}{ds} = -i \bar{a}^2 \bar{b}^* \label{eq:feom}
\eeq
%}

The bounds are $0 \leq s \leq 1$.  
If we express the phase terms of $\bar{x}$ with $\phi_x$, we obtain the following equations for the signal amplitude and phase. 
\bea 
	\frac{d|\bar{a}|}{ds} &=& |\bar{a}||\bar{b}||\bar{c}| \cos \theta \\
	\frac{d \phi_a}{ds} &=& |\bar{b}||\bar{c}| \sin \theta \\
	\theta &=& \phi_b + \phi_c - 2\phi_a +\pi/2
\eea
Thus, we can realize a PSA where only the signal with phase 0 or $\pi$ relative to the sum of pump phases is amplified. We may assume $\phi_b = \phi_c = 0$ without loss of generality. Since we are interested in above-threshold behaviour, $\bar{a}$ may be presumed real because only the real quadrature of $\bar{a}$ experiences gain. So in the following we treat $\bar{x}$ as real in Eq. (\ref{eq:feom}). 

Equation (\ref{eq:feom}) has two constants of motion: $A_b^2 = \bar{a}^2 + 2\bar{b}^2$ and $A_c^2 = \bar{a}^2 + 2\bar{c}^2$, which arise from the detailed balance in the $a + a \leftrightarrow b + c$ process.  Upon integrating, we find:

\beq
	\bar{a} = \sqrt{\frac{1 - \tanh^2(A_b A_c (s + s_0)/2)}{A_b^{-2} - A_c^{-2} \tanh^2(A_b A_c (s + s_0)/2)}} \label{eq:as}
\eeq

Given the initial conditions $\bar{a}_{in}, \bar{b}_{in}, \bar{c}_{in}$, we may invert (\ref{eq:as}) to find the constant of integration $s_0$, and then directly compute $\bar{a}_{\rm out}$.  This gives the input-output relation.  Rescaling to physical units, we define the input-output map $F[a]$ so that $a_{\rm out} = F[a_{\rm in}] e^{-\alpha L/2}$, i.e.\ $F[a]$ accounts for the PSA gain but not for its loss.

Let $a_i(m)$ represent the $i^{\rm th}$ pulse at round-trip $m$.  To calculate the field at $m+1$, the pulse passes through the nonlinear fibre and is then split in the delay line.  Defining $G_0$ as the total round-trip (power) loss, we find:

\beq
	a_i(m+1) = \frac{F[a_i(m)] \pm F[a_{i-1}(m)]}{2\sqrt{G_0}} \label{eq:map}
\eeq

Ferromagnetic interactions use a $+$ sign; antiferromagnetic interactions use a $-$ sign. 
%in this section we assume a ferromagnetic interaction.  
To account for quantum noise, we work in a truncated Wigner picture \cite{drum}, which is convenient and accurate when the threshold photon number is $\gg 1$.  This procedure adds Gaussian noise terms to (\ref{eq:map}) to maintain the commutation relations $[a_i, a_j^\dagger] = \delta_{ij}$ in the presence of cavity and coupling losses.  

Equation (\ref{eq:map}) is simulated numerically to obtain the curve shown in Fig. \ref{1d} {\bf d}. In the simulation, the pump amplitudes were turned on at $m=0$ and kept at the same values until the end of the simulation, $m=1000$. We performed the simulation for various pump amplitudes, and the defect density $n_d$ was derived from the final spin configuration for each pump amplitude.  

%%%%%%%%%%%%%analytical 
The simulations show that the evolution is a two-stage process: in the {\it growth stage}, the field is weak and pump depletion may be ignored, giving rise to exponential growth in the fields $a_i$.  This lasts for time $T_s$ and is followed by a {\it crystallization stage}, where the field saturates to one of two values: $a \rightarrow \pm a_{\rm sat}$ and domain walls form.  Over time, the domain walls attract each other and this causes smaller domains to evaporate.

%Most of the dynamics happens in the growth stage.  
With the reasonable approximation stated above, the dynamics in the growth state can be analytically solved as follows. 
%The exact result depends on the ratio of the pump amplitude to threshold: $b/b_0$.  
At the threshold, the fibre gain must be $G_0$ to compensate for loss.  Above the threshold for $a \ll b, c$, the input-output relation can be linearized to give $F[a] = G_0^{\frac{1}{2} p} a$.  Applying (\ref{eq:map}), the net effect of a single round trip is:

\beq
	a_i(t+1) = G_0^{\frac{1}{2}(p-1)} \frac{a_i(t) + a_{i-1}(t)}{2} \label{eq:grmap1}
\eeq

The linear map (\ref{eq:grmap1}) is diagonalized by going to the Fourier domain.  We then integrate the equation up to time $T_s = (p-1)^{-1} \frac{\log(N_{\rm sat})}{\log(G_0)}$, namely the time it takes to reach saturation.  
$T_s$ depends only logarithmically on the saturation photon number $N_{\rm sat}$, which is $O(10^6-10^9)$. 
%, factors of two or three are not significant, so we can estimate $N_{sat} \rightarrow b^2$, the pump energy.  
The field amplitude at saturation is:

\beq
	\tilde{a}_k(T_s) \sim \sqrt{N_{\rm sat}} e^{-\frac{T_s}{2}(\pi k/N)^2} \label{eq:akdist}
\eeq

This has a Gaussian power spectral density, which in turn gives a Gaussian autocorrelation function: $R(x) = \langle a_i a_{i+x}\rangle/\langle a_i^2\rangle = e^{-x^2/2x_0^2}$, with the autocorrelation length given by $x_0 = \sqrt{T_s/2}$.  
%Given that the $a_i(T_s)$ are normally distributed, a simple argument from probability theory shows that the defect density, which equals the jump probability $P(a_i a_{i+1} < 0)$, is given by $n_d = 1/\pi x_0$.  It follows that the average domain length must be $L_d = 1/n_d = \pi x_0$.

We analysed the crystallization stage with numerical simulations, and found that $a_i$ saturates to $\pm a_{sat}$ and as a result $a_k$ changes its form. We also found that the auto-correlation curves obtained in the simulations changed from the Gaussian to a function that was approximated by $\exp(-x/x'_0)$ with $x'_0 \simeq x_0$, and thus the auto-correlation of the DOPOs may be reasonably approximated as the spin-correlation of the  1D Ising spins. 

\end{document}